\documentclass[12pt]{article}
\usepackage{amsmath}
\usepackage{graphicx}
\begin{document}
\baselineskip=18 pt
\begin{center}
{\large{\bf A type N radiation field solution with $\Lambda<0$ in a curved space-time and closed time-like curves}}
\end{center}

\vspace{.5cm}

\begin{center}
{\bf Faizuddin Ahmed}\footnote{faizuddinahmed15@gmail.com}\\
{\it Ajmal College of Arts and Science, Dhubri-783324, Assam, India}
\end{center}

\vspace{.5cm}

\begin{abstract}

An anti-de Sitter background four-dimensional type N solution of the Einstein's field equations, is presented. The matter-energy content pure radiation field satisfies the null energy condition (NEC), and the metric is free-from curvature divergence. In addition, the metric admits a non-expanding, non-twisting and shear-free geodesic null congruence which is not covariantly constant. The space-time admits closed time-like curves which appear after a certain instant of time in a causally well-behaved manner. Finally, the physical interpretation of the solution, based on the study of the equation of the geodesics deviation, is analyzed. 

\end{abstract}

\vspace{0.3cm}


\vspace{.5cm}

\section{Introduction}

The energy-momentum tensor of radiation field \cite{Steph} is given by
\begin{equation}
T^{\mu\nu}=\rho\,k^{\mu}\,k^{\nu},\quad k^{\mu}\,k_{\mu}=0,
\label{1}
\end{equation}
where $\rho$ is the radiation energy-density and $k^{\mu}$ is the tangent vector field of geodesics null congruence. The trace-free and covariant conservation of this tensor $T^{\mu\nu}_{\,\,\,;\,\nu}=0$ implies that the radiation propagates along geodesic {\it i.e.,} 
\begin{equation}
k_{\mu;\nu}\,k^{\nu}=0.
\label{2}
\end{equation}
The cosmological constant Einstein's field equations with radiation field are given by
\begin{equation}
R_{\mu\nu}-\frac{1}{2}\,g_{\mu\nu}\,R+\Lambda\,g_{\mu\nu}=\rho\,k_{\mu}\,k_{\nu},
\label{3}
\end{equation}
where $R_{\mu\nu}$ is the Ricci tensor, $R$ is the scalar curvature, and $\Lambda$ is the cosmological constant. Taking trace of the field equations, one will get
\begin{equation}
R=4\,\Lambda.
\label{4}
\end{equation}
Substituting this into the field equations (\ref{3}), one will get
\begin{equation}
R_{\mu\nu}=\Lambda\,g_{\mu\nu}+\rho\,k_{\mu}\,k_{\nu},\quad \mu,\nu=0,1,2,3.
\label{5}
\end{equation}
In terms of traceless Ricci tensor $S_{\mu\nu}$, the field equations can be written as
\begin{equation}
S_{\mu\nu}=R_{\mu\nu}-\frac{1}{4}\,g_{\mu\nu}\,R=\rho\,k_{\mu}\,k_{\nu}.
\label{6}
\end{equation}
The type N solutions of the field equations play a fundamental role in the theory of gravitational radiation. The various algebraically special Petrov type have some interesting physical interpretations in the context of gravitational radiation. Particularly for type N space-time, it has only one repeated principal null direction (PND) of multiplicity $4$, which means that all four PNDs coincide. The non-vanishing components of the Weyl scalar is $\Psi_4$ and this corresponds to transverse gravitational wave propagate along geodesics null congruence. A complete class of non-twisting type N vacuum solutions with $\Lambda \neq 0$ was obtained in \cite{Garcia}, and a geometrically different vacuum solution with $\Lambda<0$ in \cite{Siklos}. The complete family of non-expanding type N vacuum solutions with $\Lambda \neq 0$ was further analysed and classified in \cite{Ozvath} (see also \cite{Bicak,Bicak2}). These Einstein spaces represent exact pure gravitational waves which propagate in Minkowski, de-Sitter or anti-de Sitter background (for $\Lambda=0$, $\Lambda>0$ or $\Lambda<0$, respectively). The algebraically special (type II, D, III or N, conformally flat) non-twisting and shear-free pure radiation space-times with or without cosmological constant are known in literature ({\it e.g.} \cite{Be,Be2,Haza,Podo,Podo2,Grif,Edg,Mach,Kai}) (see \cite{Steph,Podo3} for a comprehensive review). In the present article, we attempt to construct a type N radiation field solution with $\Lambda<0$. The space-time admits a non-expanding, non-twisting and shear-free geodesic null congruence which is not covariantly constant null vector (CCNV). That means, the space-time exhibit geometrically different properties than plane-fronted gravitational waves with parallel rays ({\it pp}-waves). In addition, the space-time display causality violation by admitting closed time-like curves. 

The presence of closed time-like curves (CTC) in a space-time violate the notion of causality in general relativity. Hawking proposed the Chronology Protection Conjecture \cite{Haw} to counter the appearance of closed time-like curves. However, the general proof of this Conjecture has not yet been known. Space-time with closed time-like curves cannot be discard or rule out because such space-times are the exact solutions of the field equations. On some physical backgrounds, for examples, space-time possesses a curvature singularity or does not admit a partial Cauchy surface and/or generate closed time-like curves which are come from infinity are considered nonphysical solutions. A few solutions content unrealistic/exotic matter-energy sources violating one or more energy conditions. For CTC space-time, the matter-energy sources must be realistic, that is, the stress-energy tensor must be known type of matter fields which satisfy the different energy conditions. Many known CTC space-time, for examples the traversable wormholes \cite{MTY1,MTY2}, and the warp drive models \cite{Alcu,Ever,Lobo} violate the weak energy condition (WEC), which states that $T_{\mu\nu}\,U^{\mu}\,U^{\nu}\geq 0$ for a time-like tangent vector field $U^{\mu}$, that is, the energy-density must be non-negative. The CTC space-time in (\cite{Soen}) violate the strong energy condition (SEC), which states that $(T_{\mu\nu}-\frac{1}{2}\,g_{\mu\nu}\,T)\,U^{\mu}\,U^{\nu}\geq 0$ (see details in \cite{Hawking}). There is another energy condition : the dominant energy condition which directly implies the weak energy condition and this implies the null energy condition, which states that $T_{\mu\nu}\,k^{\mu}\,k^{\nu}\geq 0$ for any null vector $k^{\mu}$. The radiation field solutions in curved space-time without cosmological constant ({\it e.g.} \cite{Sarma,Faiz1,Faiz2,Faiz3}) develops closed time-like curves. Thus if the null energy condition is satisfied the three other energy conditions are also satisfied. Therefore the null energy condition appears to be the most fundamental among all the energy conditions since it cannot be violated by the addition of a suitably large vacuum energy contribution.

The present work comprises into four section : in {\it section 2}, a four-dimensional curved space-time with negative cosmological constant and pure radiation field, is analyzed, in {\it section 3}, the physical interpretation of the solution, will be discussed, and finally conclusions in {\it section 4}.

Our conventions are : Greek indices are taking values $0,1,2,3$ and Einstein's summation convention is used. The choice of signature is $(-,+,+,+)$ and the units are chosen $c=1=8\,\pi\,G=\hbar$.

\section{A radiation field space-time with $\Lambda<0$}
 
Consider the following time-dependent metric in $(t,r,\psi,z)$ coordinates given by
\begin{equation}
ds^2=g_{rr}\,dr^2+g_{zz}\,dz^2+2\,g_{t\psi}\,dt\,d\psi+g_{\psi\psi}\,dz^2+2\,g_{z\psi}\,d\psi\,dz,
\label{7}
\end{equation}
where the different metric functions are 
\begin{eqnarray}
g_{rr}&=&\mbox{coth}^{2} (\sqrt{\frac{-\Lambda}{3}}\,r),\quad \Lambda<0,\nonumber\\
g_{\psi\psi}&=&-\sinh t\,\sinh^{2} (\sqrt{\frac{-\Lambda}{3}}\,r),\nonumber\\
g_{t\psi}&=&-\frac{1}{2}\,\cosh t\,\sinh^{2} (\sqrt{\frac{-\Lambda}{3}}\,r),
\label{8}
\end{eqnarray}
\begin{eqnarray}
g_{z\psi}&=&\beta_0\,z\,\sinh^{2} (\sqrt{\frac{-\Lambda}{3}}\,r),\nonumber\\
g_{zz}&=&\sinh^{2} (\sqrt{\frac{-\Lambda}{3}}\,r)\nonumber,
\end{eqnarray}
with $\beta_0>0$ is a real number. The ranges of the coordinates are 
\begin{equation}
-\infty < t < \infty,\quad 0 \leq r < \infty,\quad -\infty < z < \infty,
\label{9}
\end{equation}
and the coordinate $\psi$ is chosen periodic. The validity of imposing periodicity on $\psi$ coordinate generally follows from the regularity condition on the axis. A space-time admitting an axial Killing vector $\eta^{\mu}$, parameterized by a $2\,\pi$-periodic coordinate $\psi$ is regular on the rotation axis (a set of fixed points of $\eta^{\mu}$) if and only if the following condition holds :
\begin{equation}
\frac{(\nabla_{\mu}{\boldsymbol{X}})\,(\nabla^{\mu}{\boldsymbol{X}})}{4\,\boldsymbol{X}}\rightarrow 1,
\label{10}
\end{equation}
where the limit corresponds to the rotation axis \cite{Steph}. In our case, we find l. h. s. corresponds to the rotation axis is $(-1-\beta_0^{2}\,z^2\,\mbox{csch} t)$ in the region $t<0$ where, $\psi$ coordinate is spacelike. Thus one can identify $\psi \sim \psi+\psi_0$ where, $\psi_0\sim (2\,\pi-\delta)<2\,\pi$ and $\delta$ is the deficit angle. There are cosmic string or conical singularity exist on the non-regular axis. In the region $t>0$, the coordinate $\psi$ becomes time-like and the space-time generate closed time-like curves which we shall discuss later in this article. 

The determinant of the metric tensor $g_{\mu\nu}$ given by
\begin{equation}
det\;g=-\frac{1}{4}\,\cosh^{2} {(\sqrt{\frac{-\Lambda}{3}}\,r)}\,\sinh^{4} (\sqrt{\frac{-\Lambda}{3}}\,r)\,\cosh^2 t,
\label{11}
\end{equation}
degenerates at $r=0$.

The non-zero components of the Ricci tensor $R_{\mu\nu}$ are
\begin{eqnarray}
R_{r\,r}&=&\Lambda\,\mbox{coth}^{2} (\sqrt{\frac{-\Lambda}{3}}\,r),\nonumber\\
R_{\psi\psi}&=&\beta_0-\Lambda\,\sinh t\,\sinh^{2} (\sqrt{\frac{-\Lambda}{3}}\,r),\nonumber\\
R_{z\psi}&=&\Lambda\,\beta_0\,z\,\sinh^{2} (\sqrt{\frac{-\Lambda}{3}}\,r),\nonumber\\
R_{t\psi}&=&-\frac{\Lambda}{2}\,\cosh t\,\sinh^{2} (\sqrt{\frac{-\Lambda}{3}}\,r),\nonumber\\
R_{zz}&=&\Lambda\,\sinh^{2} (\sqrt{\frac{-\Lambda}{3}}\,r).
\label{Ricci-tensor}
\end{eqnarray}
The non-zero components of the Weyl tensor $C_{\mu\nu\rho\sigma}$ and the Riemann tensor $R_{\mu\nu\rho\sigma}$ are
\begin{eqnarray}
C_{1212}&=&-\frac{\beta_0}{2}\,\mbox{coth}^2 (\sqrt{\frac{-\Lambda}{3}}\,r),\quad C_{2323}=\frac{\beta_0}{2}\,\sinh^2 (\sqrt{\frac{-\Lambda}{3}}\,r),\nonumber\\
R_{1212}&=&\frac{-\Lambda}{3}\,\cosh^2 (\sqrt{\frac{-\Lambda}{3}}\,r)\,\sinh t,\quad R_{0202}=\frac{-\Lambda}{12}\,\sinh^4 (\sqrt{\frac{-\Lambda}{3}}\,r)\,\cosh^2 t,\nonumber\\
R_{1313}&=&\frac{\Lambda}{3}\,\cosh^2 (\sqrt{\frac{-\Lambda}{3}}\,r),\quad R_{1213}=\frac{\Lambda}{3}\,\beta_0\,z\,\cosh^2 (\sqrt{\frac{-\Lambda}{3}}\,r),\nonumber\\
R_{2323}&=&\sinh^2 (\sqrt{\frac{-\Lambda}{3}}\,r)\,[\beta_0-\frac{\Lambda}{3}\,\sinh^2 (\sqrt{\frac{-\Lambda}{3}}\,r)\,(\beta_0^{2}\,z^2+\sinh t)],\nonumber\\
R_{0232}&=&\beta_0\,z\,R_{0332},\quad R_{0121}=\frac{-\Lambda}{6}\,\cosh^2 (\sqrt{\frac{-\Lambda}{3}}\,r)\,\cosh t,\nonumber\\
R_{0332}&=&\frac{\Lambda}{6}\,\sinh^4 (\sqrt{\frac{-\Lambda}{3}}\,r)\,\cosh t.
\label{weyl-tensor}
\end{eqnarray}

For the presented time-dependent metric, there are following Killing vector fields 
\begin{eqnarray}
\xi_{1}^{\mu}&=&\eta^{\mu}=\left(0,0,1,0\right),\nonumber\\
\xi_{2}^{\mu}&=&\left(-2\,\sqrt{\frac{-\Lambda}{3}}\,\mbox{tanh} t, \mbox{tanh} (\sqrt{\frac{-\Lambda}{3}}\,r), 0, -\sqrt{\frac{-\Lambda}{3}}\,z\right),
\label{12}
\end{eqnarray}
\begin{eqnarray}
\xi_{3}^{\mu}&=&\left(e^{-\psi}\,\mbox{sech} t,0,0,0\right),\nonumber\\
\xi_{4}^{\mu}&=&e^{\frac{1}{2}\,(-1+\sqrt{1-4\,\beta_0})\,\psi}\left(z\,(-1+\sqrt{1-4\,\beta_0}+2\,\beta_0)\,\mbox{sech} t,0,0,1\right),\nonumber\\
\xi_{5}^{\mu}&=&e^{\frac{1}{2}\,(-1-\sqrt{1-4\,\beta_0})\,\psi}\left(z\,(-1-\sqrt{1-4\,\beta_0}+2\,\beta_0)\,\mbox{sech} t,0,0,1\right)\nonumber.
\end{eqnarray}
The space-time (\ref{7})--(\ref{8}) satisfy the field equations (\ref{5}) provided the source energy-density
\begin{equation}
\rho=\beta_0>0,\quad \Lambda<0
\label{13}
\end{equation}
for the following null vector of the metric
\begin{equation}
k_{\mu}=(0,0,1,0)=\delta^{\psi}_{\mu}.
\label{14}
\end{equation}
Thus the radiation energy-density ($\rho$) which is a constant satisfy the null energy condition (NEC) since the metric function $g^{\psi\psi}=0$. Noted that the null vector (\ref{14}) satisfies the condition (\ref{2}) with 
\begin{equation}
\boldsymbol{\Theta}=\frac{1}{2}\,k^{\mu}_{\,;\,\mu}=0,\,\,\boldsymbol{\omega}^2=\frac{1}{2}\,{k_{[\mu\,;\,\nu]}}\,k^{\mu\,;\,\nu}=0,\,\,\boldsymbol{|\sigma|}^2=\frac{1}{2}\,{k_{(\mu\,;\,\nu)}}\,k^{\mu\,;\,\nu}-\boldsymbol{\Theta}^2=0.
\label{24}
\end{equation}
The quantities $\boldsymbol{\Theta}$, $\boldsymbol{\omega}$ and $\boldsymbol{\sigma}$ are called the {\it expansion}, the {\it twist} and the {\it shear}, respectively. Hence this null vector field can be considered as the tangent vector field of geodesic null congruence the radiation propagates along. But this null vector field is not covariantly constant null vector (CCNV), that is, $k_{\mu;\nu}\neq 0$. Therefore the studied space-time exhibit geometrically different properties than the famous known {\it pp}-waves space-time.

To show the studied space-time is free-from curvature divergence, we have calculated the following curvature invariant constructed from the Riemann tensor as:
\begin{eqnarray}
R^{\mu\nu}\,R_{\mu\nu}&=&4\,\Lambda^2,\,\,R^{\mu\nu\rho\sigma}\,R_{\mu\nu\rho\sigma}=\frac{8}{3}\,\Lambda^2,\,\,R_{\mu\nu\rho\sigma}\,R^{\rho\sigma\lambda\tau}\,R^{\mu\nu}_{\,\,\,\lambda\tau}=\frac{16}{9}\,\Lambda^3,\nonumber\\
R^{,\mu}\,R_{,\mu}&=&0,\,R^{\mu\nu;\tau}\,R_{\mu\nu;\tau}=0,\,R^{\mu\nu\rho\sigma;\tau}\,R_{\mu\nu\rho\sigma;\tau}=0.
\label{15}
\end{eqnarray}
Similarly, we have calculated the following curvature invariant constructed from the Weyl tensor as: 
\begin{eqnarray}
\centering
I_{1}&=&C_{\mu\nu\rho\sigma}\,C^{\mu\nu\rho\sigma}=0,\quad I_{2}=C_{\mu\nu\rho\sigma}\,C^{*\,\mu\nu\rho\sigma}=0,\nonumber\\
I_{3}&=&C_{\mu\nu\rho\sigma;\tau}\,C^{\mu\nu\rho\sigma;\tau}=0,\quad I_{4}=C_{\mu\nu\rho\sigma;\tau}\,C^{*\,\mu\nu\rho\sigma;\tau}=0,
\label{16}
\end{eqnarray}
where $C_{\mu\nu\rho\sigma}$ is the Weyl tensor, and $C^{*}_{\mu\nu\rho\sigma}$ its dual. 

From the above analysis, it is clear that the curvature invariants constructed from the Riemann tensor and the Weyl tensor do not blow up which guaranteed that the studied space-time is free-from curvature divergence.

Now we discuss closed time-like curves of the space-time which appear after a certain instant of time. Consider an azimuthal closed curves $\gamma$ defined by $t=t_0$, $r=r_0$, and $z=z_0$ where, $t_0,r_0>0,z_0$ are constants and $\psi$ is periodic, $\psi\sim \psi+\psi_0$ where, $\psi_0>0$. From the metric (\ref{7}), we get
\begin{equation}
ds^2=-\sinh t_0\,\sinh^{2} (\sqrt{\frac{-\Lambda}{3}}\,r_0)\,d\psi^2.
\label{17}
\end{equation}
There are time-like curves provided $ds^2<0$ for $t=t_0>0$, spacelike provided $ds^2>0$ for $t=t_0<0$, and null curve  $ds^2=0$ for $t=t_0=0$. Therefore the closed curves defined by $(t,r,\psi,z) \sim (t_{0},r_{0},\psi+\psi_0,z_{0})$ being time-like in the region $t=t_0>0$, formed closed time-like curves. Noted the Gott's time-machine space-time generated closed time-like curves by imposing one of the coordinate $\psi$ is periodic identifying $\psi\sim \psi+\psi_0$ with period $\psi_0<2\,\pi$ \cite{Gott} (see also \cite{Ori,Leva}). These time-like closed curves evolve from an initial spacelike $t=const<0$ hypersurface \cite{Ori,Leva}. We find from metric (\ref{7}) that the metric component $g^{00}$ is given by
\begin{equation}
g^{00}=4\,\mbox{csch}^{2} (\sqrt{\frac{-\Lambda}{3}}\,r)\,\mbox{sech}^2 t\,\left(\beta_0^{2}\,z^2+\sinh t\right).
\label{18}
\end{equation}
Now we have chosen the constant $z-planes$ defined by $z=z_0$, where $z_0$, a constant equal to zero. Therefore, from (\ref{18}) we get
\begin{equation}
g^{00}=\frac{4\,\sinh t}{\sinh^{2} (\sqrt{\frac{-\Lambda}{3}}\,r)\,\cosh^2 t}.
\label{19}
\end{equation}
A hypersurface $t=const=t_0$ is spacelike ($r=r_0>0$) provided $g^{00}<0$ for $t<0$, and time-like provided $g^{00}>0$ for $t>0$. Therefore the spacelike $t=const=t_0<0$ hypersurface can be chosen as initial conditions over which the initial data may specified. There is a Cauchy horizon at $t=t_0=0$ for any such spacelike $t=const=t_0<0$ hypersurface. The null curve at $t=t_0=0$ serve as the Chronology horizon (since $g^{00}=0$) which divided the space-time a chronal region without CTC to a non-chronal region with CTC. Hence, the space-time evolves from an initial spacelike hypersurface in a causally well behaved manner, up to a moment, {\it i.e.}, a null hypersurface $t=t_0=0$, and the formation of CTC takes place from causally well behaved initial conditions in the $z=const$-planes.

\section{Further analysis of the space-time}

In this section, we first classify the presented space-time according to the Petrov classification scheme, and its physical interpretation will be the subsequent part.

\subsection{Classification of the metric}

We construct a set of tetrad vectors $({\bf k,l,m,{\bar m}})$ for the presented metric. These are given by
\begin{eqnarray}
k_{\mu}&=&(0,0,1,0),\nonumber\\
l_{\mu}&=&\frac{1}{2}\,\sinh^2 (\sqrt{\frac{-\Lambda}{3}}\,r)\,(\cosh t, 0, \sinh t, -2\,\beta_0\,z),\nonumber\\
m_{\mu}&=&\frac{1}{\sqrt{2}}\,\left(0, \mbox{coth} (\sqrt{\frac{-\Lambda}{3}}\,r), 0, i\,\sinh (\sqrt{\frac{-\Lambda}{3}}\,r)\right ),\nonumber\\
{\bar m}_{\mu}&=&\frac{1}{\sqrt{2}}\,\left(0,\mbox{coth} (\sqrt{\frac{-\Lambda}{3}}\,r), 0, -i\,\sinh (\sqrt{\frac{-\Lambda}{3}}\,r)\right ).
\label{20}
\end{eqnarray}
The set of tetrad vectors are such that the metric tensor for the line element (\ref{7})-(\ref{8}) is
\begin{equation}
g_{\mu \nu}=-k_{\mu}\,l_{\nu}-l_{\mu}\,k_{\nu}+m_{\mu}\,\bar{m}_{\nu}+\bar{m}_{\mu}\,m_{\nu},
\label{21}
\end{equation}
where the tetrad vectors are null and orthogonal except $k_{\mu}\,l^{\mu}=-1$ and $m_{\mu}\,{\bar m}^{\mu}=1$.

Using the above tetrad vectors, we calculate the five Weyl scalars and these are
\begin{equation}
\Psi_0=\Psi_1=0=\Psi_2=\Psi_3,\quad \Psi_{4}=-\frac{\beta_0}{2}.
\label{22}
\end{equation}
In addition, the Weyl tensor $C_{\mu\nu\rho\sigma}$ satisfies the following Bel criteria,
\begin{equation}
C_{\mu\nu\rho\sigma}\,k^{\sigma}=0.
\label{23}
\end{equation}
Thus the metric (\ref{7})-(\ref{8}) is of type N in the Petrov classification scheme. One can calculate the Newmann-Penrose spin coefficients \cite{Steph} for the presented metric. These are given by
\begin{eqnarray}
\tau&=&-\pi=\alpha=-\sqrt{\frac{-\Lambda}{6}},\quad \gamma=-\frac{1}{2},\quad \nu=-\frac{i\,\beta_0\,z}{\sqrt{2}}\,\sinh (\sqrt{\frac{-\Lambda}{3}}\,r),\nonumber\\ \kappa&=&\rho=\sigma=\epsilon=\mu=\lambda=\beta=0,
\label{25}
\end{eqnarray}
where the symbols are same as in \cite{Steph}. Thus the repeated principal null direction $\boldsymbol{k}$ aligned with the radiative direction is geodesic and shear-free.  

The Riemann and Ricci tensor satisfies the following relation  
\begin{eqnarray}
R_{\mu\nu\rho\sigma}\,k^{\sigma}&=&\frac{\Lambda}{3}\,(g_{\mu\rho}\,k_{\nu}-g_{\nu\rho}\,k_{\mu}),\quad R_{\mu\nu}\,k^{\mu}=\Lambda\,k_{\nu},\nonumber\\
&&R_{\mu\nu\rho\sigma}\,k^{\rho}\,k^{\sigma}=0=R_{\mu\nu}\,k^{\mu}\,k^{\nu}.
\label{aa}
\end{eqnarray}
The complex scalar quantities $\Phi_{AB}={\bar \Phi}_{AB}$, $A,B=0,1,2$ associated with the trace-free Ricci tensor $S_{\mu\nu}$ are
\begin{eqnarray}
\Phi_{00}&=&\frac{1}{2}\,S_{\mu\nu}\,k^{\mu}\,k^{\nu}=0,\quad \Phi_{01}=\frac{1}{2}\,S_{\mu\nu}\,k^{\mu}\,m^{\nu}=0,\nonumber\\
\Phi_{02}&=&\frac{1}{2}\,S_{\mu\nu}\,m^{\mu}\,m^{\nu}=0,\quad \Phi_{11}=\frac{1}{2}\,S_{\mu\nu}\,(k^{\mu}\,l^{\nu}+m^{\mu}\,{\bar m}^{\nu})=0,\nonumber\\
\Phi_{12}&=&\frac{1}{2}\,S_{\mu\nu}\,l^{\mu}\,m^{\nu}=0,\quad \Phi_{22}=\frac{1}{2}\,S_{\mu\nu}\,l^{\mu}\,l^{\nu}=\frac{\beta_0}{2}.
\label{tetrad-ricci}
\end{eqnarray}
An orthonormal tetrad frame ${\bf e}_{(a)}=\{{\bf e}_{(0)},{\bf e}_{(1)},{\bf e}_{(2)},{\bf e}_{(3)}\}$ in terms of null tetrad vectors (\ref{20}) can be express as
\begin{equation}
{\bf k}=\frac{1}{\sqrt{2}}({\bf e}_{(0)}+{\bf e}_{(2)}),\quad {\bf l}=\frac{1}{\sqrt{2}}({\bf e}_{(0)}-{\bf e}_{(2)}),\quad {\bf m}=\frac{1}{\sqrt{2}}({\bf e}_{(1)}+i\,{\bf e}_{(3)}),
\label{26}
\end{equation}
where ${\bf e}_{(0)}\cdot{\bf e}_{(0)}=-1$ and ${\bf e}_{(i)}\cdot{\bf e}_{(j)}=\delta_{ij}$.

\subsection{The relative motion of free test particles}

In order to analyze the effects of the gravitational field and matter field of the above solution, we used the technique adopted in \cite{Pirani1,Pirani2,Szek1,Szek2}. The equation of geodesic deviation frame \cite{Bicak2,Podo2} are given by
\begin{equation}
\frac{D^{2}Z^{\mu}}{d\tau^2}=-R^{\mu}_{\,\nu\rho\sigma}\,u^{\nu}\,Z^{\rho}\,u^{\sigma},\quad {\bf u}={\bf e}_{(0)},
\label{ss}
\end{equation}
where ${\bf u}\cdot{\bf u}=-1$ is the four-velocity of a free test particle (observer), and $Z^{\mu}(\tau)$ is the displacement vector connecting two neighbouring free test particles. The equations of geodesic deviation in terms of orthonormal tetrad frame (\ref{26}) are
\begin{equation}
\ddot{Z}^{(i)}=-R^{(i)}_{\,(0)(j)(0)}\,Z^{(j)},\quad i,j=1,2,3,
\label{27}
\end{equation}
where $Z^{(i)}\equiv e^{(i)}_{\mu}\,Z^{\mu}$ are frame components of the displacement vector and $\ddot{Z}^{(i)}\equiv e^{(i)}_{\mu}\,\frac{D^{2}Z^{\mu}}{d\tau^2}$ are relative accelerations. Here we set $Z^{(0)}=0$ so that all test particles are synchronized by the proper time.

From the standard definition of the Weyl tensor using (\ref{6}) one will get
\begin{equation}
R_{(i)(0)(j)(0)}=C_{(i)(0)(j)(0)}+\frac{1}{2}\,(\delta_{ij}\,S_{(0)(0)}-S_{(i)(j)})-\frac{\Lambda}{3}\,\delta_{ij}.
\label{28}
\end{equation}
The non-vanishing Weyl scalars are given (\ref{22}) so that 
\begin{equation}
C_{(1)(0)(1)(0)}=-\frac{\beta_0}{4}=C_{(1)(2)(1)(2)},\quad C_{(3)(0)(3)(0)}=\frac{\beta_0}{4}=C_{(2)(3)(2)(3)},
\label{29}
\end{equation}
and rest are all vanish.

The equations of geodesic deviation (\ref{27}) using (\ref{28})--(\ref{29}) are
\begin{eqnarray}
\ddot{Z}^{(1)}&=&-R^{(1)}_{\,(0)(j)(0)}\,Z^{(j)}=\frac{\Lambda}{3}\,Z^{(1)},\nonumber\\
\ddot{Z}^{(2)}&=&-R^{(2)}_{\,(0)(j)(0)}\,Z^{(j)}=\frac{\Lambda}{3}\,Z^{(2)},\\
\label{30}
\ddot{Z}^{(3)}&=&-R^{(3)}_{\,(0)(j)(0)}\,Z^{(j)}=(\frac{\Lambda}{3}-\frac{\beta_0}{2})\,Z^{(3)}\nonumber,
\end{eqnarray}
with the solutions
\begin{eqnarray}
Z^{(1)}(\tau)&=&A_1\,\cos (\sqrt{\frac{-\Lambda}{3}}\,\tau)+B_1\,\sin (\sqrt{\frac{-\Lambda}{3}}\,\tau),\nonumber\\
Z^{(2)}(\tau)&=&A_2\,\cos (\sqrt{\frac{-\Lambda}{3}}\,\tau)+B_2\,\sin (\sqrt{\frac{-\Lambda}{3}}\,\tau),\\
\label{31}
Z^{(3)}(\tau)&=&A_3\,\cos (\sqrt{\frac{-\Lambda}{3}+\frac{\beta_0}{2}}\,\tau)+B_3\,\sin (\sqrt{\frac{-\Lambda}{3}+\frac{\beta_0}{2}}\,\tau),\nonumber
\end{eqnarray}
where $A_i, B_i, i=1,2,3$ are arbitrary constants and 
\begin{equation}
S_{(0)(0)}=\frac{\beta_0}{2}=S_{(2)(2)},\quad S_{(1)(1)}=0=S_{(3)(3)}.
\label{32}
\end{equation}
If one takes $\Psi_4=0$ which implies $\beta_0=0$, the equations of geodesic deviation reduces to
\begin{equation}
\ddot{Z}^{(i)}=\frac{\Lambda}{3}\,Z^{(i)}
\label{33}
\end{equation}
with the solutions
\begin{equation}
Z^{(i)}=A_i\,\cos (\sqrt{\frac{-\Lambda}{3}}\,\tau)+B_i\,\sin (\sqrt{\frac{-\Lambda}{3}}\,\tau).
\label{34}
\end{equation}
All test particles move isotropically one with respect to the other which explains the resulting isotropic motions. Thus the term proportional to $\Lambda$ in (\ref{31}) represent the influence of anti-de Sitter (AdS) backgrounds. Clearly, the relative motions of nearby test particles depend on : the Ricci scalar $R$ (or cosmological constant $\Lambda$) responsible for overall background isotropic motions ; the influence of local free gravitational field $\Psi_{4}$, and the stress-energy tensor $T_{(a)(b)}$ terms describing interaction of matter-content both of their amplitudes depend on the real number $\beta_0$.

\section{Conclusions}

We presented a four-dimensional radiation field type N solution of the Einstein's field equations with negative cosmological constant ($\Lambda<0$). The presence of a negative cosmological constant implies that the background space is not asymptotically flat. The studied metric is non-diverging ($\rho=-(\boldsymbol{\omega}+i\,\boldsymbol{\Theta})=0$), has a shear-free ($\boldsymbol{\sigma}=0$) geodesic null vector field which is considered the principal null direction aligned with radiative direction. This null vector field is not a covariantly constant vector field, that means, the rays of transverse gravitational wave are not parallel and therefore the studied metric is geometrically different from the known {\it pp}-waves. Furthermore, we shown the space-time admits closed time-like curves which appear after a certain instant of time. These time-like closed curves evolve from an initial spacelike $t=const<0$ hypersurface in a causally well behaved manner in the $z=const$-planes. A reasonable physical interpretation of a space-time is possible if one investigates the equation of geodesic deviation in a suitable frame. We investigated the physical interpretation of the presented solution, based on the equation of the geodesic deviation in an orthonormal tetrad frame ${\bf e}_{(a)}$. It was demonstrated that, this space-time can be understood as exact transverse gravitational waves propagating in an everywhere curved anti-de Sitter Universe, and the matter-energy sources radiation field which affect the relative motion of the free-test particles.

\end{document}